\documentclass[pra, twocolumn]{revtex4-1}
\usepackage[english]{babel}
\usepackage[latin1]{inputenc}
\usepackage{hyperref}
\usepackage{amsmath, amssymb, amsfonts, mathrsfs}
\usepackage{graphicx}
\graphicspath{{images/}}
\usepackage{xcolor}
\usepackage{exscale}
\usepackage{graphicx}
\usepackage{amsmath}
\usepackage{latexsym}
\usepackage{amsfonts}
\usepackage{amssymb}

\DeclareMathOperator{\Tr}{Tr}
\DeclareMathOperator{\Id}{ {\bf 1}}

\def\pmx{\begin{pmatrix}}
\def\emx{\end{pmatrix}}

\newcommand{\ket}[1]{|#1\rangle}
\newcommand{\bra}[1]{ \langle #1 \,  |}

\begin{document} 

\title{Quantum Cloning by Cellular Automata}

\author{G.M. D'Ariano}\email{dariano@unipv.it}
\author{C. Macchiavello}
\author{M. Rossi}\email{matteo.rossi@unipv.it}

\affiliation{Dipartimento di Fisica and INFN-Sezione di Pavia,
Via Bassi 6, 27100 Pavia, Italy}

\begin{abstract}
  We introduce a quantum cellular automaton that achieves approximate phase-covariant cloning of qubits.
  The automaton is optimized for $1\rightarrow 2N$ economical cloning.  The use of the automaton for
  cloning allows us to exploit different foliations for improving the performance with given resources.
\end{abstract}

\date{\today}

\maketitle

\section{introduction}

Quantum cellular automata (QCAs) have attracted considerable interest in recent years
\cite{Werner1,Werner2}, due to their versatility in tackling several problems in quantum physics. Quantum
automata describing single particles correspond to the so-called quantum random walks \cite{rwalk},
whose probability distributions can be simulated in with an optical setup \cite{Silberhorn,Mataloni}.
Solid-state and atom-optics systems, such as spin-chains, optical lattices, or ion chains, can be viewed
as implementations of QCAs, though in a Hamiltonian description.  Recently QCAs have also been
considered as a model of quantum field theory at the Planck scale \cite{dpla,bdt}. 

In this scenario, general coordinate transformations correspond to
foliations, such as those introduced in Ref.  \cite{Hardy} for operational structures, e.g. the digital equivalent of the relativistic boost is given by a uniform foliation
over the automaton \cite{darx1}. Besides the link with fundamental research, the possibility of foliations makes the QCA particularly interesting also for implementing quantum
information tasks. In this  paper we will explore such a potentiality for the case of quantum
cloning as a sample protocol. We will show how economical cloning can be implemented by a quantum
automaton, and how the foliations can be optimized and exploited for improving the efficiency of the
protocol.

The work is organized as follows. In Sec. \ref{pre} we remind some concepts related to either phase-covariant cloning and QCAs, in Sec. \ref{phase} we report and explain the main achieved results concerning quantum cloning by QCAs, and we eventually summarized them in Sec. \ref{conc}.

\section{preliminaries}\label{pre}

It is well known that quantum cloning of nonorthogonal states violates unitarity \cite{yuen} or
linearity \cite{Zurek} of quantum theory, and it is equivalent to the impossibility of measuring the
wave-function of a single system \cite{yuendar}. However, one can achieve quantum cloning
approximately, for a given prior distribution over input quantum states.  For uniform Haar
distribution of pure states the optimal protocol has been derived in Ref. \cite{Werner}, whereas for
equatorial states it has been given in Refs. \cite{Cinchetti,cm}.  In the present  paper we consider
specifically this second protocol, corresponding to clone the two-dimensional equatorial states
\begin{equation}\label{equatorial}
\ket{\phi}=\frac{1}{\sqrt 2}(\ket{0}+e^{i\phi}\ket{1}).
\end{equation}
The cloning is phase covariant in the sense that its performance is independent of $\phi$, i.e.
the fidelity is the same for all states $\ket{\phi}$. 
For certain numbers of input and output copies it was shown that the optimal 
fidelity can be achieved by a 
transformation acting only on the input and 
blank qubits, without extra ancillas \cite{Durt,Bus}. 
Since these transformations act only on the minimal number of qubits they
are called ``economical''.  The unitary operation $U_{pcc}$ realising the 
optimal $1\rightarrow 2$
economical phase-covariant cloning is given by \cite{Durt}
\begin{align}\label{1->2}
&U_{pcc}\ket{0}\ket{0}=\ket{0}\ket{0},\\ \nonumber
&U_{pcc}\ket{1}\ket{0}=\frac{1}{\sqrt 2}(\ket{0}\ket{1}+\ket{1}\ket{0}),
\end{align}
where the first qubit is the one we want to clone, while the second is the blank qubit initialised
to input state $\ket{0}$.  In Ref. \cite{Bus} the
economical map performing the optimal $N\rightarrow M$ phase-covariant 
cloning for equatorial states
of dimension $d$ is explicitly derived for $M=kd+N$ with integer $k$.

\begin{figure}
\includegraphics[width=1\linewidth]{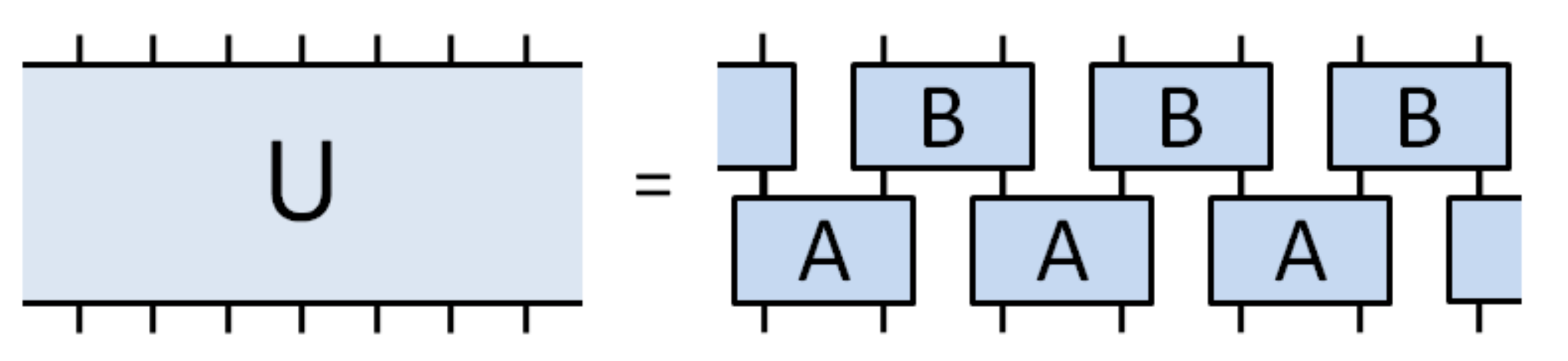}
\caption{Realisation of one-dimensional quantum cellular automaton with a structure composed
  of two layers of gates $A$ and $B$. This is the most general one-dimensional automaton with next-nearest neighbour interacting minimal cells.}\label{QCA}
\end{figure}

In order to analyse a QCA implementation of the economical quantum cloning, we now recall the reader some properties of QCAs we are considering here. Our automaton is one-dimensional, and a single time-step corresponds to a unitary shift-invariant
transformation achieved by two arrays of
identical two-qubit gates in the two-layer Margolus scheme
\cite{Werner1} reported in Fig. \ref{QCA}. Notice that this is the most general one-dimensional automaton with next-nearest neighbour interacting minimal cells.
Due to the locality of interactions, information about a qubit cannot be
transmitted faster than two-systems per time step, and this corresponds to 
the cell (qubit) ``light-cone'' made of cells that are causally connected 
to the first. Every event outside the cone has no
chance to be influenced by what happened in the first cell, thus the 
quantum computation of the
evolution of localized qubits is finite for finite numbers of time-steps.

\begin{figure}
\includegraphics[width=.8\linewidth]{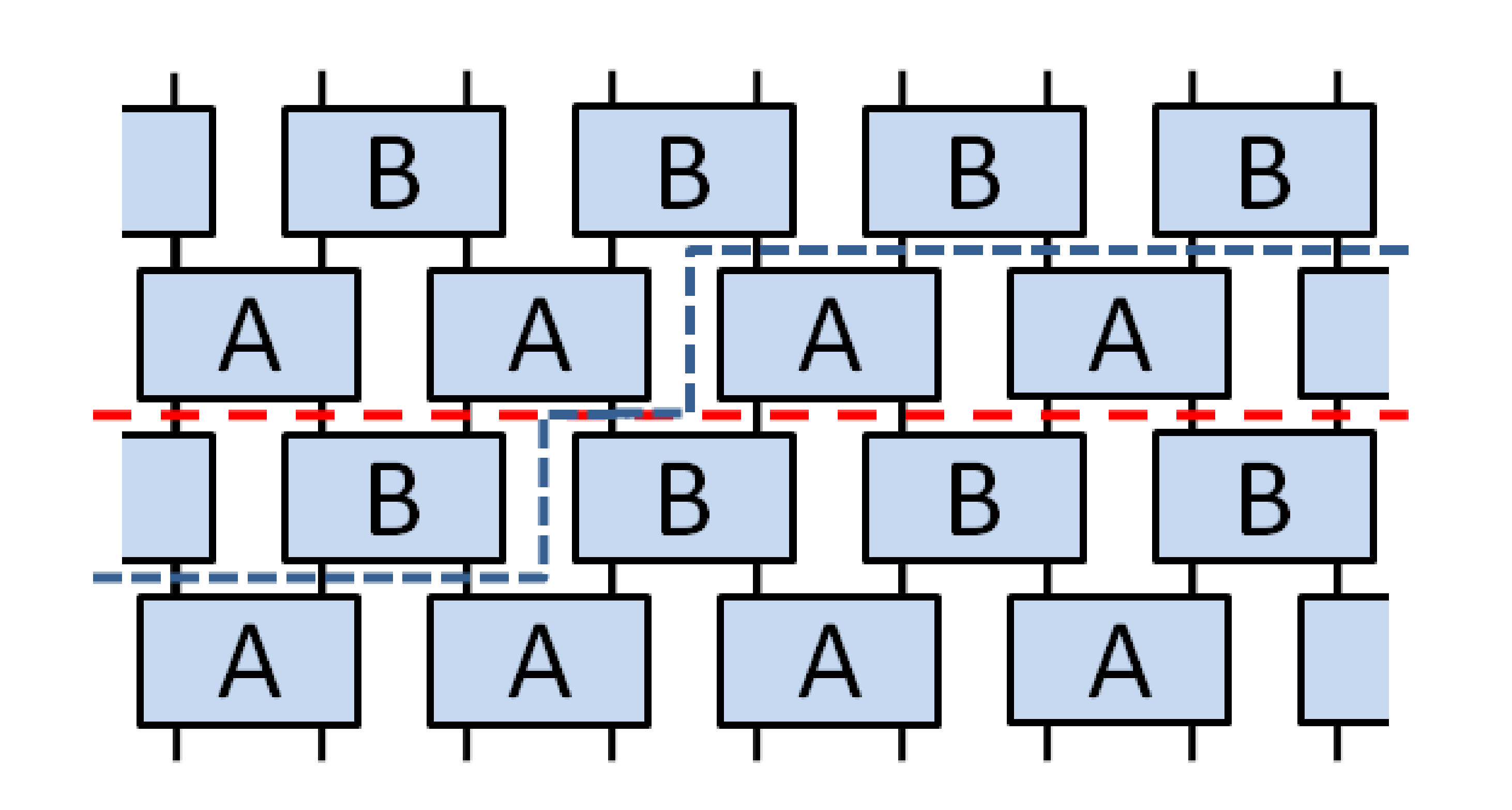}
\caption{(Colors online) Foliations over the automaton. Two leafs of two different uniform
  foliations are depicted with dash-lines in different colors (the complete foliation is obtained
  upon repeating vertically the leaf).  The systems along each leaf are taken as simultaneous.  The
  red ``cut'' is usually referred to as the rest-frame foliation.}
\label{foliations}
\end{figure}

We now remind the concept of {\em foliation} on the gate-structure of the QCA \cite{darx1}.  Usually
in a quantum circuit--drawn from the bottom to the top as the direction of input output--one
considers all gates with the same horizontal coordinate as simultaneous transformations. A foliation
on the circuit corresponds to stretching the wires (namely without changing the connections), and
considering as simultaneous all the gates that lie on the same horizontal line after the stretching.
Such horizontal line can be regarded as a {\em leaf} of the foliation on the circuit before the
stretching transformation. Therefore, a foliation corresponds to a specific choice of simultaneity
of transformations (the ``events''), namely it represents an observer or a reference frame. Examples
of different foliations are given in Fig.  \ref{foliations}.  Upon considering the quantum state at
a specific leaf as the state at a given time (at the output of simultaneous gates), different
foliations correspond to different state evolutions achieved with the same circuit.  Therefore, in
practice we can achieve a specific state belonging to one of the different evolutions, by simply
cutting the circuit along a leaf, and tapping the quantum state from the resulting output wires (the
operation of ``stretching'' wires should be achieved by remembering that by convention the wires
represent identical evolutions, not ``free'' evolutions).

\section{Phase-covariant cloning by QCAs}\label{phase}

\begin{figure}
\includegraphics[width=.6\linewidth]{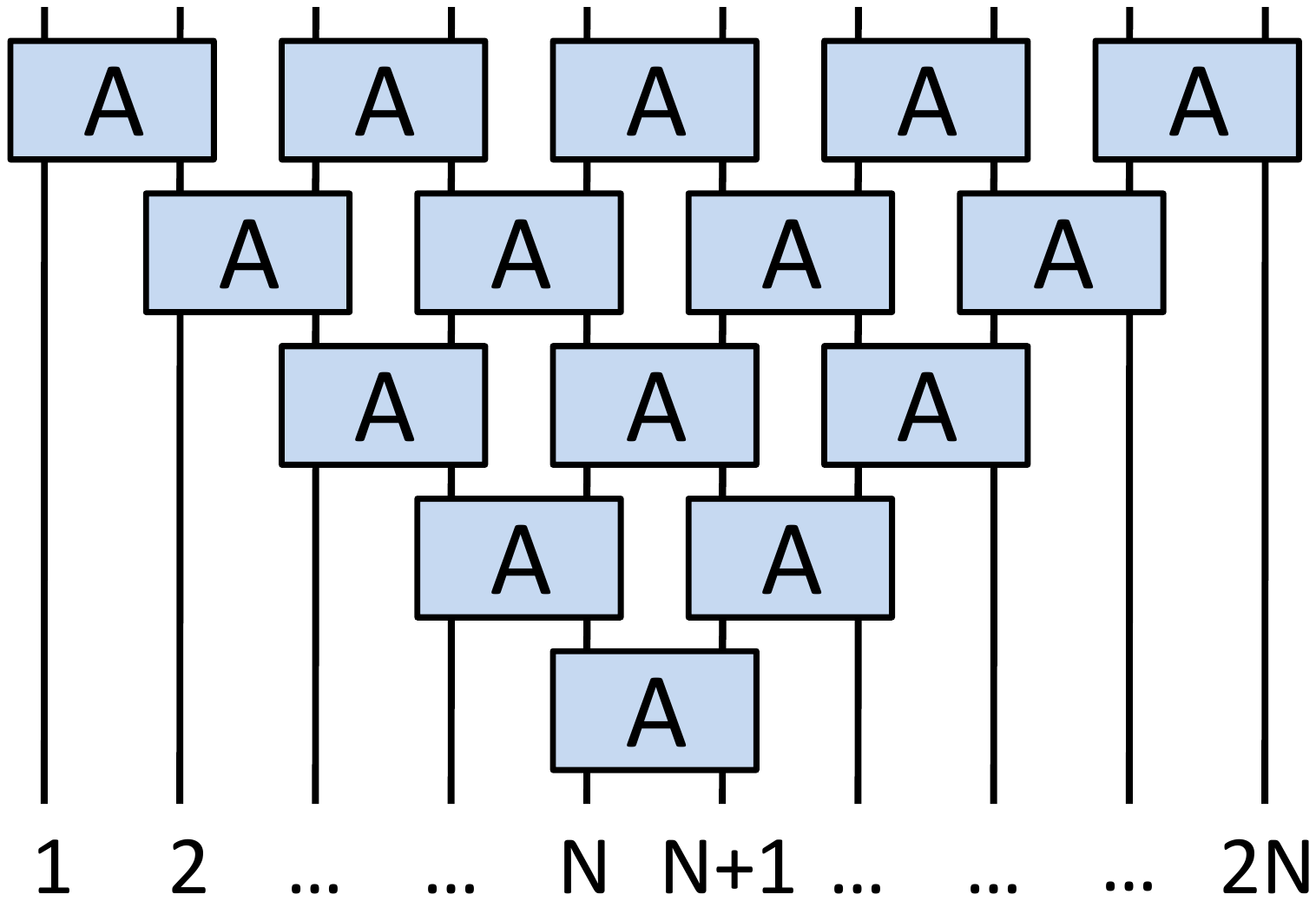}
\caption{Cone of gates which contribute to the phase-covariant 
cloning, given the input state $\ket{\phi}$ at site $N$.}
\label{cone}
\end{figure}

We will now show how to perform a $1\rightarrow 2N$ phase-covariant cloning of the equatorial states
\eqref{equatorial} with a QCA of $N$ layers, with all gates identical, performing the unitary
transformation denoted by $A$, acting on two qubits.  Due to causality, we can restrict our
treatment to the light-cone centered in the state to be cloned $\ket{\phi}$ and initialise all blank
qubits to $\ket{0}$, as shown in Fig. \ref{cone}. By requiring phase-covariance for the cloning
transformation, the unitary operator $A$ must commute with every transformation of the form
$P_\chi\otimes P_\chi$, where $P_\chi$ is the general phase-shift operator $P_\chi=\exp
[\frac{i}{2}(\Id-\sigma_z)\chi]$ for a single qubit, with $\sigma_z$ the Pauli matrix along $z$.
Therefore, we impose the condition
\begin{equation}\label{comm}
[A,P_\chi\otimes P_\chi]=0,\text{ }\forall \chi\;.
\end{equation}
This implies that the matrix $A$ must be of the form $A=\text{diag}(1,V,1)$, where $V$ is a $2\times
2$ unitary matrix.
Notice that the transformation $A$ then acts non-trivially only on the 
subspace spanned by the two states $\{\ket{01},\ket{10}\}$ and it is 
completely specified by fixing $V$. 

In order to derive the optimal cloning transformation based on this kind of QCA we will now maximise
the average single-site fidelity of the $2N$-qubits output state with respect to the unitary
operator $A$.  In order to achieve this, we write the initial state of $2N$ qubits in the following
compact form
\begin{equation}
\ket{\Psi_0}=\frac{1}{\sqrt 2}(\ket{\Omega}+e^{i\phi}\ket{N}),
\end{equation}
where we define $\ket{\Omega}=\ket{0\dots0}$ as the ``vacuum state'' with all qubits in the $0$
state, and $\ket{k}=\ket{0\dots01_k0\dots0}$ as the state with the qubit up in the position $k$, and
all other qubits in the state down. Without loss of generality in the above notation the qubit to be
cloned is supposed to be placed at position $N$ and it is initially in the state $\ket{\phi}$. Since
the gate $A$ preserves the number of qubits up \cite{excitation}, the evolved state through each layer
will belong to the Hilbert space spanned by the vacuum state and the $2N$ states with one qubit up.
The whole dynamics of the QCA can then be fully described in a Hilbert space of
dimension $2N+1$. The output state can thus be  generally written as
\begin{equation}\label{final}
\ket{\Psi_{2N}}=\frac{1}{\sqrt 2}(\ket{\Omega}+e^{i\phi}
\sum_{k=1}^{2N}\alpha_k\ket{k}),
\end{equation}
where the amplitudes $\alpha_k$ of the excited states depend only on the explicit form of the gate
$A$.

The reduced density matrix $\rho_k$ of the qubit at site $k$ can then be derived from the output
state \eqref{final} as
\begin{align}
\rho_k  =&\Tr_{\bar k}[\ket{\Psi_{2N}}\bra{\Psi_{2N}}] \\ \nonumber
	= &\frac{1}{2}\Big[ \big(1+\sum_{j\neq k}|\alpha_j|^2\big)\ket{0}\bra{0} +e^{-i\phi}\alpha_k^*\ket{0}\bra{1} \\ \nonumber
	&+ e^{i\phi}\alpha_k\ket{1}\bra{0}+|\alpha_k|^2\ket{1}\bra{1} \Big],
\end{align}
where $\Tr_{\bar k}$ denotes the trace on all qubits except qubit $k$. 
The local fidelity of the qubit at site $k$ with respect to the input state
$\ket{\phi}$ then takes the simple form
\begin{equation}\label{fidelity}
F_k=\bra{\phi}\rho_k\ket{\phi}=\frac{1}{2}(1+\text{Re}\{\alpha_k\}).
\end{equation}
As we can see, $F_k$ depends only on the amplitude $\alpha_k$ of the state with a single qubit-up
exactly at $k$.
Since the gate $A$ is generally not invariant under exchange of the two qubits, the fidelities at
different sites will be in general different. We will then consider the average fidelity $\bar
F=\frac{1}{2N}\sum_{k=1}^{2N}F_k$ as figure of merit to evaluate the performance of the
phase-covariant cloning implemented by QCA.  Notice that the whole procedure corresponds to a
unitary transformation on the $2N$-qubit system, without introducing auxiliary systems, namely it is
an economical cloning transformation.

The calculation of the amplitudes $\alpha_k$ was performed numerically by updating at each layer the
coefficients of the state \eqref{final}.  Notice that the amplitude of layer $j$ and site $k$
influences only the amplitudes of the subsequent layer $j+1$ and sites either $k-1,k$ or $k,k+1$,
depending on whether the state $|1\rangle$ enters in the right or left wire of $A$. The action of $A$ on
the qubits $j$ and $j+1$ is given by
\begin{equation}\label{iter1}
A(j,j+1)\ket{k}=
\begin{cases} 
v_{22}\ket{j}+v_{12}\ket{j+1} & \text{if }k=j \\ 
v_{21}\ket{j}+v_{11}\ket{j+1} & \text{if }k=j+1 \\
\ket{k} & \text{otherwise}\;,
\end{cases}
\end{equation}
where $v_{ij}$ are the entries of the operator $V$ in the basis $\{\ket{01},\ket{10}\}$.  Notice
that the vacuum state $\ket{\Omega}$ is invariant under the action of $A$.  The iteration of Eq.
(\ref{iter1}) for each layer then leads to the amplitudes of the output state \eqref{final}.

\subsection{Performances in the Rest Frame}

\begin{figure}
\includegraphics[width=1\linewidth]{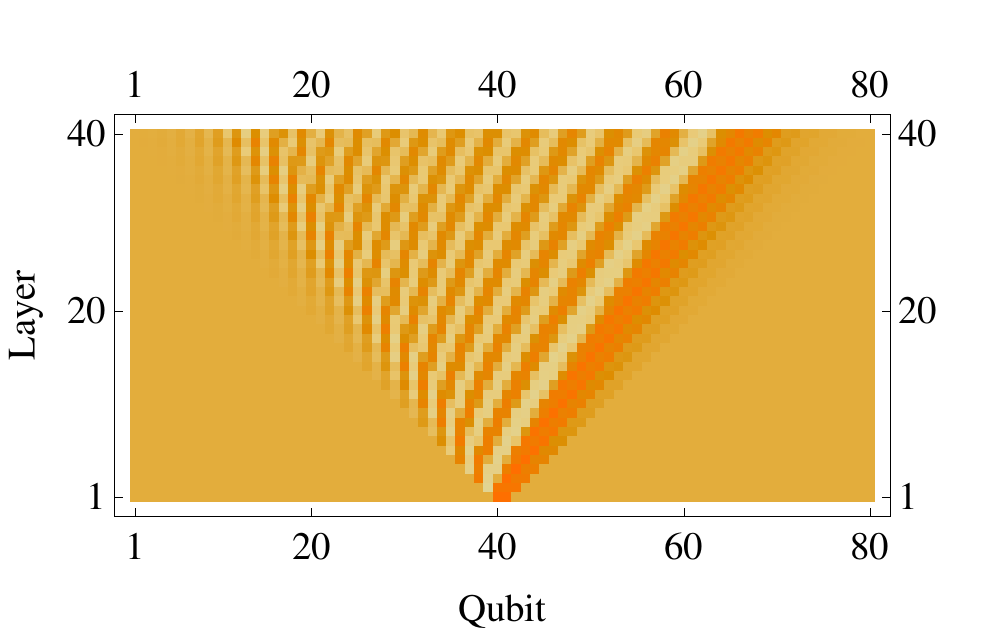}
\caption{(Colors online). Chromatic map of the local fidelities in terms of the considered qubit
  and the layer. The orange colour is brighter for increasing local fidelity. The simulation
  involves a number of layers $N=40$, while the total number of qubits is $2N$, since it doubles at
  each layer.}
\label{Mat_F}
\end{figure}

\begin{figure}
\includegraphics[width=1\linewidth]{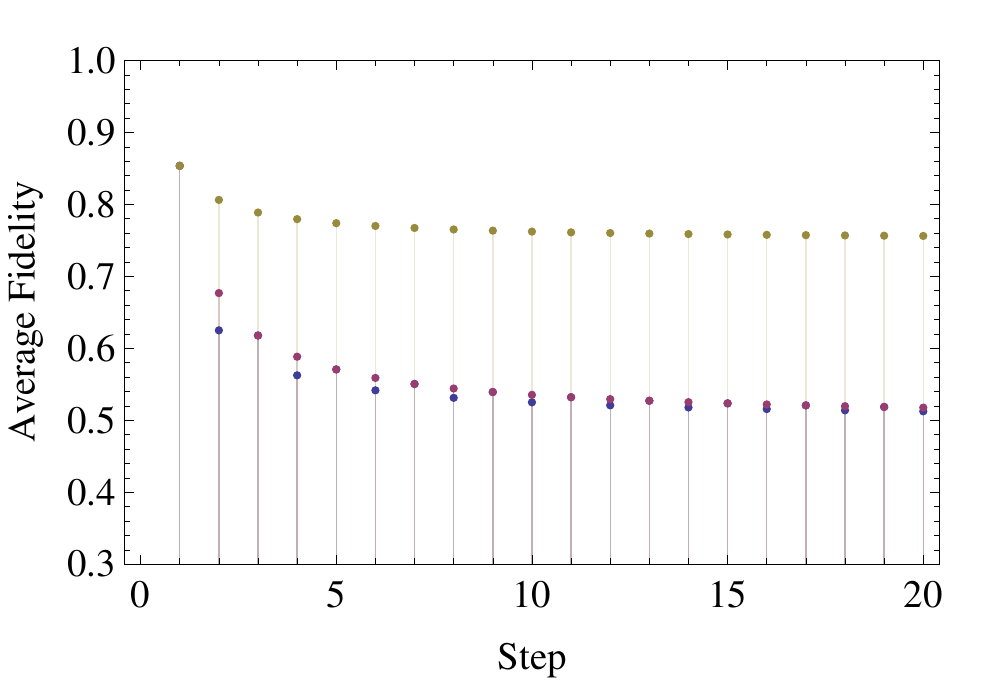}
\caption{(Colors online) The average $1\to 2N$ phase-covariant cloning fidelity achieved by the QCA  in the rest frame (see Fig. \ref{foliations}). Comparison with the optimal economical  phase-covariant cloning in Ref. \cite{cm}.  The purple dots represents the QCA cloning optimized  over the unitary gate $A$. The blue dots correspond to the use of gate $A$ achieving the optimal $1\to 2$ cloning $U_{pcc}$ in Eq. (\ref{1->2}). The yellow dots represent the unrestricted optimal economical cloning.}
\label{AF_all}
\end{figure}

As a first explicit example we will consider a QCA employing the optimal $1\rightarrow 2$
phase-covariant cloning \eqref{1->2}. In this case the gate $A$ must implement the unitary
transformation \eqref{1->2}.  The non-trivial part $V$ of gate $A$ can then be chosen to be
\begin{equation}\label{U}
V=\frac{1}{\sqrt 2}
\begin{pmatrix}
1&1\\
-1&1\\
\end{pmatrix},
\end{equation}
where all coefficients are real. 
The corresponding local fidelities at every layer are reported in Fig. 
\ref{Mat_F}.
As we can see, the figure exhibits fringes of light and dark colour.  Moreover, the light-cone
defined by causality can be clearly seen: outside this cone no information about the initial state
can arrive, thus every system has the same fidelity of $1/2$. Notice that there is a sort of line,
approaching the right top corner, along which the fidelity is quite high.  Regarding the local
fidelities of the final states, they are in general quite different from each other, and can vary
very quickly even between two neighbouring qubits. The average fidelity is reported in Fig.
\ref{AF_all} as a function of the number of layers.
Notice that the average fidelity of the optimal economical phase-covariant quantum cloning (without
the constraint of automaton structure) approaches the value $3/4$ for a large number of output
copies \cite{cm}.

In order to improve the average fidelity we have then maximised it with respect to the four
parameters defining the unitary operator $V$.  Numerical results achieved up to $N=20$ show that the
optimal cloning performed in this case is not much better than the one given by the iteration of
\eqref{U}. Eventually, the latter turns out to be outperformed only when the number of layers composing the
automaton is odd, as shown in Fig. \ref{AF_all}.

Further numerical results show that no gain can be achieved if the automaton is composed of layers
of two different gates $A$ and $B$. Actually, in this case it surprisingly turns out that the
optimal choice corresponds to $B=A$, namely we do not exceed the average fidelity obtained by
employing a single type of gate. As a result, since all one-dimensional QCA with next-nearest neighbour
interacting cells with two qubits can be implemented by a two-layered structure, we have then
derived the optimal phase-covariant cloning transformation achievable by the minimal one-dimensional QCA.

\begin{figure}[t!]
\includegraphics[width=1\linewidth]{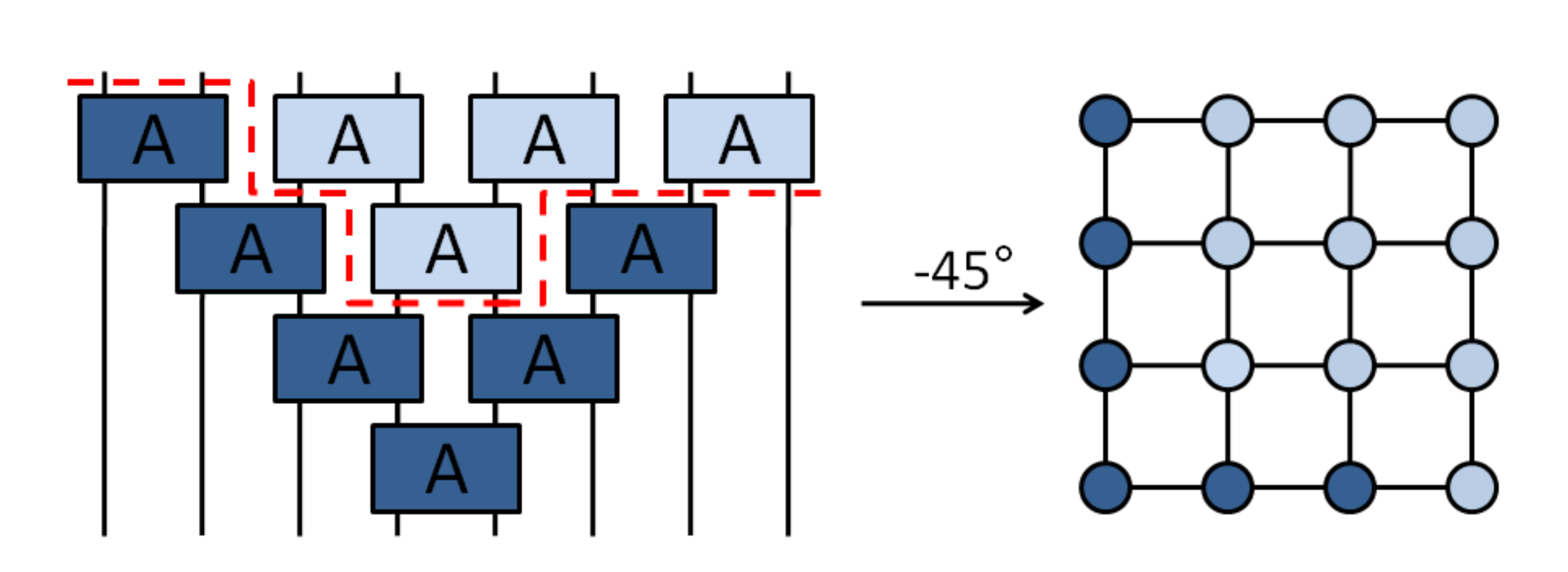}
\caption{Illustration of the classification of foliations. A possible foliation with
  $M=6$ gates is given. From the correspondence between the gates lying under the ``cut'' and the
  rotated dots on the right, we identify this foliation with the partition $\{4,1,1\}$.}
\label{partition}
\end{figure}

\subsection{Performances Exploiting Different Foliations}

We will now show that the average fidelity in the case of a single-gate automaton can be improved by considering different foliations.
Suppose that we are given a fixed number $M$ of identical gates $A$ to implement a QCA.  We are then
allowed to place the gates in any way such that the causal structure of the considered automaton is
not violated.  Which is the configuration, i.e. the foliation, that performs the optimal
phase-covariant cloning for fixed $M$?  In this framework we have to maximise not only over the
parameters that define $V$ but also over all possible foliations.  Thus, the $M$ fixed gates play
the role of computational resources, and the optimality is then defined in terms of both the
parameters characterising the single gate $A$ and the disposition of the gates in the network.  As a
first example, suppose that we are given $M=3$ gates.  In this case there are $3$ inequivalent
foliations: one for the rest frame (see Fig. \ref{foliations}), and two along the straight lines defining the light-cone.  As
expected, for increasing $M$ the counting of foliations becomes more complicated
and the problem is how to choose and efficiently investigate each possible foliation. It turns out
that the problem of identifying all possible foliations of a QCA of the form illustrated in Fig.
\ref{cone} for a fixed number of gates $M$ is related to the partitions of the integer number $M$
itself (by partition we mean a way of writing $M$ as a sum of positive integers, a well known
concept in number theory \cite{combinatorics}).  
\begin{table}[t!]
\begin{tabular}{ccccc}\hline
$M$(Layers) & $\bar F_{\text{rest}}$ & $\bar F$ & Optimal foliation \\ \hline
$1(1)$ & $0.853$ & $0.853$ & $\{1\}$ \\ 
$3(2)$ & $0.676$ & $0.693$ & $\{3\}$ \\ 
$6(3)$ & $0.617$ & $0.679$ & $\{2,2,2\}$ \\ 
$10(4)$ & $0.588$ & $0.670$ & $\{4,3,3\}$ \\  
$15(5)$ & $0.570$ & $0.653$ & $\{4,4,4,3\}$ \\ 
$21(6)$ & $0.558$ & $0.614$ & $\;\;\{4,3,2,2,2,2,2,2,2\}\;$ \\ 
$28(7)$ & $0.550$ & $0.603$ & $\{6,6,6,5,5\}$ \\ \hline
\end{tabular}\caption{Results of the maximisation over foliations 
up to $M=28$, i.e. QCA composed of up to $7$ layers.}\label{tab}
\end{table}
Two sums that differ only in the
order of their addends are considered to be the same partition.
For instance the partitions of $M=3$
are exactly $3$ and given by $\{3\}$, $\{2,1\}$, and $\{1,1,1\}$, while the partitions of $M=6$,
corresponding to a 3-layer setting in the rest frame, are $11$ and given by $\{6\}$, $\{5,1\}$,
$\{4,2\}$, $\{4,1,1\}$, $\{3,3\}$, $\{3,2,1\}$, $\{3,1,1,1\}$, $\{2,2,2\}$,$\{2,2,1,1\}$,
$\{2,1,1,1,1\}$, and $\{1,1,1,1,1,1\}$.  
The link between foliations and partitions is illustrated
in Fig. \ref{partition}, which shows how partitions can be exploited to identify foliations.  For a
fixed number of gates $M$ the number of foliations is then automatically fixed and each foliation
corresponds to a single partition.  
The correspondence is obtained as follows. Each addend
represents the number of gates along parallel diagonal lines, starting from the vertex of the
light-cone, as shown in Fig. \ref{partition} for the particular case of $M=6$.

Based on this correspondence, we can investigate the performance of the phase-covariant cloning as
follows. For any fixed foliation, we first maximise the average fidelity with respect to the  four parameters of the unitary $V$, defining the gate $A$.
Then we choose the highest average fidelity  that we have obtained by varying the foliation.  We worked out this procedure
numerically for $M=1,3,6,10,15,21,28$, i.e. the number of gates composing the QCA with
$N=1,2,3,4,5,6,7$ layers, respectively. Our results are shown in Table \ref{tab}, where the
maximisation in the rest frame is also reported for comparison. As we can see, exploiting different
foliations leads to a substantial improvement of the average fidelity.

\section{conclusions}\label{conc}

In summary, we have introduced a way of achieving quantum cloning through QCAs. We have derived
the optimal automaton achieving economical phase covariant cloning for qubits. We have shown how the
fidelity of cloning can be improved by varying the foliation over the QCA, with fixed total number
of gates used. By developing an efficient method to identify and classify foliations by means of
number theory, we have optimised the performance of the QCA phase covariant cloning for a given
fixed number of gates, and obtained in this way the most efficient foliation.


\section*{Acknowledgements} 

M.R. gratefully acknowledges an enlightening discussion with Davide A. Costanzo.


\begin{thebibliography}{99}
\bibitem{excitation} It is easy to show that, since 
$[P_\chi\otimes P_\chi,\sum_k \sigma_{z,k}]=0$ for all $\chi$, the gate $A$ 
commutes with the operator $\sum_k\sigma_{z,k}$ and thus the number of 
qubits up is preserved during the evolution.
\end{thebibliography}
\end{document}